# Unresolved problems in superconductivity of CaC$_6$

I. I. Mazin[*], L. Boeri[a], O.V. Dolgov[a], A.A. Golubov[b], G.B. Bachelet[c], M. Giantomassi[d], and O.K.Andersen[a]

[*]Center for Computational Materials Science, Naval Research Laboratory, Washington, DC 20375

[a]Max Planck Institute fur Festkorperforschung, Stuttgart, Germany

[b]Twente University, Faculty of Science and Technology, Enschede, Netherlands

[c]INFM and Dipartimento di Fisica, La Sapienza, Roma, Italy

[d]Unité PCPM, Université Catholique de Louvain, Louvain-la-Neuve, Belgium

**Abstract**

We discuss the current status of the theory of the "high-temperature" superconductivity in intercalated graphites YbC$_6$ and CaC$_6$. We emphasize that while the general picture of conventional, phonon-driven superconductivity has already emerged and is generally accepted, there are still interesting problems with this picture, such as weak-coupling regime inferred from specific heat suggesting coupling exclusively with high-energy carbon phonons coming in direct contradiction with the isotope effect measurements suggesting coupling exclusively with the low-energy intercalant modes. At the same time, the first principle calculations, while explaining T$_c$, contradict both of the experiments above by predicting *equal* coupling with both groups of phonons.

**1. Historical Introduction**

The fact that graphite, a zero-gap semiconductor, becomes superconducting upon intercalation, has been known for some decades already[1]. However, till recently[2] the critical temperature was not higher than a few Kelvins. Therefore, the discovery of superconductivity in YbC$_6$ at 6.5 K, and, a year later, in CaC$_6$[2] at 11.5 K, came as a substantial surprise.

Initially, the fact that Yb often exhibits intermediate valent states inspired speculations of a superconductivity mediated by valence fluctuations. It was found, however[3], by both experimental and theoretical investigations, that Yb remains divalent in this compound and its f-electrons play little role in the electronic structure at the Fermi level. Upon the discovery of superconductivity in CaC$_6$ Csanyi et al[4], comparing band structures of superconducting and non-superconducting intercalated graphites, discovered a correlation between the occupancy of the so-called interlayer band and the appearance (and, to some extent, critical temperature) of superconductivity.

This undisputable correlation led Csanyi et al to a logical conjecture that this band (its character and formation will be discussed below) must play some important role in superconductivity. However, their next conjecture, that this band does not couple to phonons, was not really based on any computational or experimental fact and turned out to be incorrect[5-7]. This fact to a large extent renders moot the discussion initiated by the exciting suggestion[4] that the superconductivity in AC$_6$ (A=Yb,Ca) is of electronic origin, but it is worth noting that there are even more general reasons to believe that neither of the two proposed electronic mechanisms, Ginzburg's "sandwich" superconductivity and 2D acoustic plasmon mechanism, can be operative in these compounds. We will discuss this briefly later in the paper and refer the reader for details to Ref. [5].

The next step in unraveling the mechanism of superconductivity in AC$_6$ was taken by one of us in Ref. [5], where it was pointed out that the square root of the mass ratio of the intercalants, Yb and Ca, is just 15% larger than the ratio of the critical temperatures. At that moment it had been reported that the crystal structure of the two com-pounds was identical. This, together with the established valency 2 of Yb in $YbC_6$, led to the conclusion[5], that the electronic structures of the two compounds are extremely similar, and therefore the difference in their critical temperatures must be mostly due to the difference in the phonon frequency of the intercalant atoms (which, incidentally, was in line with the trend discovered by Csanyi *et al*, that filling of the interlayer band correlates with superconductivity). Thus, a prediction was made that the isotope effect on Ca should not be much reduced from its ideal value of 0.5 (15% reduction being a natural rule-of-thumb guess).

However, Emery *et al* turned the next page in the story of $CaC_6$,[9] by showing that its crystal structure is actually different from that of $YbC_6$, not much but enough to render Mazin's estimate[5] inaccurate. Indeed, first principle calculations of Mauri and Calandra[6] and of Boeri et al[7] showed that the intercalant modes do not *dominate* the electron phonon coupling, but contribute to it about as much as graphite modes.

These calculations predicted a moderately strong coupling ($\lambda \approx 0.85$). Solving the Eliashberg equations with the calculated $\alpha^2 F(\omega)$, with a reasonable $\mu=0.12$, gives the experimental $T_c$. While the conclusion that superconductivity in $AC_6$ is conventional, s-wave, phonon-mediated, has found full support in numerous experiments, as discussed below, there are still alarming quantitative discrepancies between the theory and the experiment, as well as among different experiments.

## 2. Electrons and phonons in $CaC_6$

As discussed below, the main feature that distinguishes the superconducting intercalated graphites is the appearance at the Fermi level of a 3-dimensional nearly-free-electron band (NFEB) that some authors (e.g., Ref. [4]) call a free electron band, and others (e.g., Ref. [6]) an intercalant band, yet others an interlayer band (e.g., Ref.[8] and refs therein). The former refer to the fact that one can remove Ca from the system and still identify the NFEB, but much higher in energy than in $CaC_6$. The latter argue that one can remove *carbon* instead, leaving a somewhat expanded along 111 metallic Ca and still see the same band! Of course, the root of the controversy is in the fact that pure Ca is a nearly-free-electron metal and its sp bands are rather similar to the free electron bands (Figs. 1,2)

From Fig.1 one can see that the NFEB is present in both cases. From Fig 2 it is obvious that there are nevertheless important differences: In $CaC_6$, calcium $d(z^2-1)$ orbitals participate in formation of this band, as well as carbon p(z). Most importantly, Fig.2 clearly shows that the the NFEB electrons are localized in the intercalant plane (even without an intercalant!), and therefore the NFEB should be rather sensible to any charge redistribution in this plane (think intercalant phonons), which it indeed is, as was first pointed out in Ref. [5] and confirmed in Refs.[6][7][15].

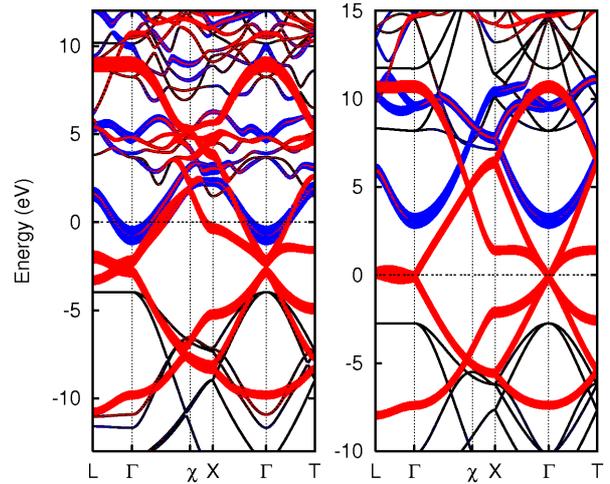

Fig. 1 Left panel: Band structure of $CaC_6$, with the NFEB emphasized in blue, and carbon $\pi$ bands in red. Right panel: the same for a non-intercalated graphite.

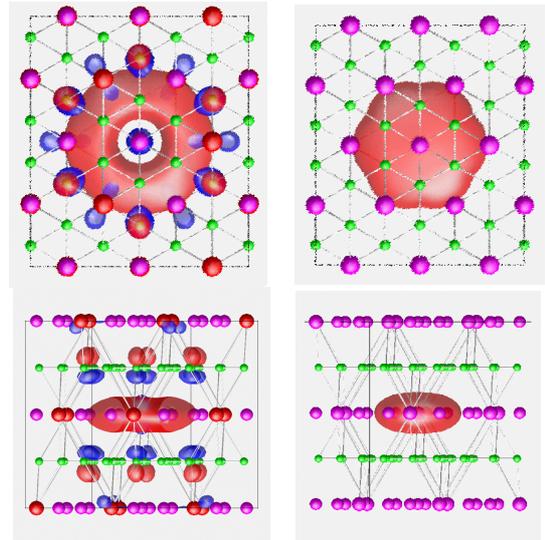

Fig. 2 Isocontours of the amplitude of the Wannier function (red: positive, blue: negative) corresponding to the NFEB. Left panel: $CaC_6$, Right panel: $C_6$.

Actual linear response calculations[6][7] have produced an interesting and consistent picture: There are three distinctive groups of modes, one at $\omega \approx 10$ meV, another at $\omega \approx 60$ meV, and the third at $\omega \approx 170$ meV, contributing, respectively, $\approx 0.4$, $\approx 0.35$, and $\approx 0.10$ to the total coupling constant $\lambda$ (Fig.3) These three groups are mainly composed of the Ca, out-of plane and in-plane C vibrations, respectively. Note that the lowest group has the frequency, in Kelvin units, of 100-120 K, while the energy of the most efficient pairing phonons is[10] $2\pi T_c=72$ K. Thus these phonons are quite close to the borderline of the applicability of the Eliashberg equations, as discussed in regards to ultrasoft modes in Ref. [11].

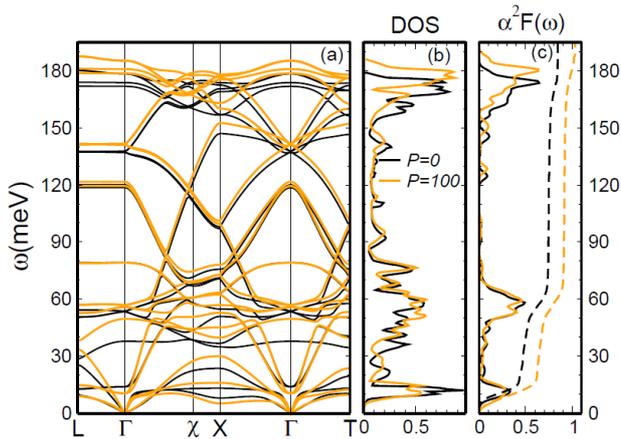

Fig. 3. (a) Calculated phonon dispersion and (b) density of states of $CaC_6$ for P = 0 and P = 100 kbar, (c) Eliashberg function and frequency-dependent electron-phonon coupling constant $\lambda(\omega)=2\int_{x<\omega}\alpha^2F(x)dx/x$, from Ref.[7].

Fig. 4 shows the calculated Fermi surface sheets of $CaC_6$. These cannot be separated into a Fermi surface associated with the NFEB and another with the carbon bands, although it is visually clear that some parts of each Fermi surface originate in one set of bands and some others in the other. It is also clear from the Figure that on average the electronic structure is fairly isotropic, thanks to the NFEB. Indeed, for $YbC_6$ the calculated plasma frequencies[5] are $\omega_{ab}$=6.2 eV and $\omega_c$=3.6 eV, and for $CaC_6$ we have computed in the same way $\omega_{ab}$=6.6 eV and $\omega_c$=3.5 eV (the corresponding average Fermi velocities are $2.4\times10^8$ and $1.2\times10^8$ cm/sec). Our final comment in this section is that up to now the coupling with the Ca phonons has been discussed only in terms of the linear harmonic response. Softness of the Ca in-plane modes and the very fact that electronically similar $CaC_6$ and $YbC_6$ assume different stacking sequences for the intercalant layers suggest that anharmonic effects, as well as nonlinear coupling (two-phonon exchange) may be operative in this system, as they possibly are in $MgB_2$[12].

Next we discuss briefly various models that have been discussed in connection with superconductivity in $AC_6$.

*Ginzburg's sandwich.* This model calls for a layered structure that consists of highly-conductive ("metallic") layers and highly polarizable ("semiconducting") layers. The idea is that the "metallic" electrons polarize the "semiconducting" layer and this dynamic polarization creates a net attractive interaction.

The key elements of this model are high polarizability and small characteristic energy for the "semiconducting" layer. The latter condition is necessary for suppressing the direct Coulomb repulsion via the Tolmachev logarithm. Unfortunately, the fact that exactly in the superconducting graphites the NFEB crosses the Fermi level renders both carbon and intercalant layers metallic thus excluding the possibility of "sandwich" superconductivity.

*Acoustic plasmons.* Just as the previous model, the idea of superconductivity due to exchange of acoustic plasmons is about 40 years old, and not dissimilar to the former. In this model, electronic systems that provide the carrier and polarization are intertwined. The same conditions apply: high polarizability, and low characteristic frequency, and one other condition appears: that of stability against a charge density wave formation. Note that the criterion for an attractive pairing interaction in the plasmon model differs from that of a CDW instability only by vertex corrections, which usually severely restricts the admissible parameter space in model calculations.

Two types of acoustic plasmons have been discussed in the literature. One refers to the Goldstone mode that appears in any system with charge carriers of vastly different masses (cf. acoustic phonons). This model is clearly inapplicable to $AC_6$, where only a very moderate electronic mass variation exists. The other model applies to a strongly 2D system, where the Fermi velocity in one direction is much smaller than in the others. Since the plasma frequency can be calculated from the latter and the density of states N(0) as $m\omega_p^2=4\pi<Nv_F^2>_{FS}$, a small Fermi velocity in one direction may lead to a very small plasma frequency in the same directions and in principle one may think about superconductivity mediated by such plasmons. Unfortunately, as mentioned above, the partial filling of the NFEB ensures that the plasma frequency along the c direction is of the order of several eV, that is, 2D acoustic phonons are not present in the system.

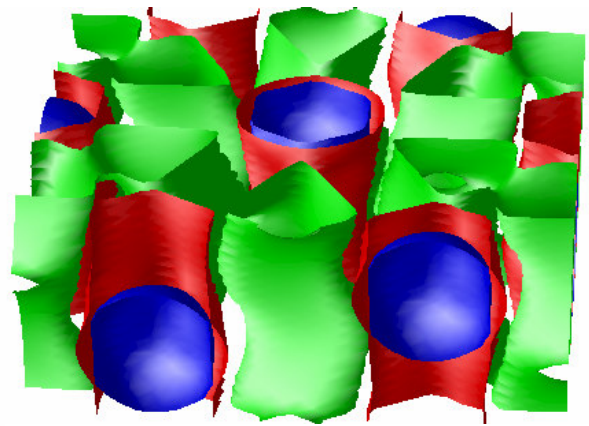

Fig. 4. Fermi surface of $CaC_6$. Note that NFEB contributes to two different Fermi surfaces, blue and red.

*Multiband superconductivity.* Although currently the interest to multigap superconductivity is high, thanks to $MgB_2$, the idea of a two-gap superconductivity in intercalated graphites is very old[13]. The problem here is that, as Fig. 4 shows, in this case one can speak, at best, about anisotropic superconductivity with an order parameter radically changing over one and the same Fermi surface sheet. While not impossible, it makes a two-gap superconductivity less likely. Indeed, experimental evidence is mounting in favor of one isotropic order parameter[14][15][16].

### 3. Unsolved problems

At first glance, it may seem that superconductivity in $AC_6$ is fully explained in terms of the conventional, standard, isotropic BCS theory. Unfortunately, this is not the case. Below we list the questions that still await answers:

*Critical field.* As pointed out in Ref. [15][18], $H_{c2}(T)$ remains linear down to 1 K, which contradicts the standard BCS model. In principle, very strong coupling can render a linear $H_{c2}(T)$ [11]; however, we have calculated $H_{c2}(T)$ using the LDA $\alpha^2F(\omega)$ and found that sublinear behavior starts already at T<4 K. A more likely explanation has to do with the Fermi velocity anisotropy, which can lead to a linear $H_{c2}$ even in a cubic superconductor like $K_3C_{60}$.[19] A good rule of thumb is that if half of the Fermi surface has a velocity twice larger than the other half, $H_{c2}(T)$ is linear [*e.g.,* according to the cubic formula $H_{c2}(0)/H_{c2}^{BCS}(0)= \langle v_F^2 \rangle / \langle \exp(\ln v_F^2) \rangle$]. However, this explanation needs a quantitative verification.

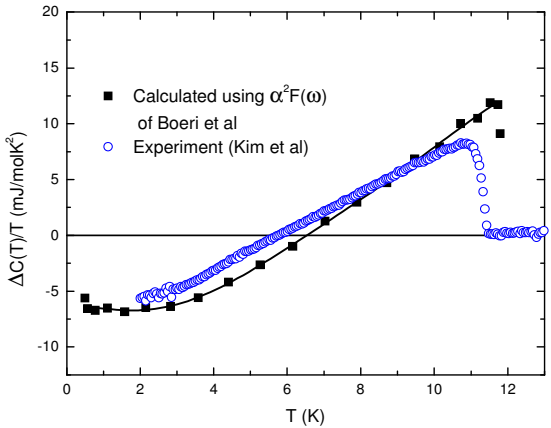

Fig. 5. Deviation of the calculated (strong coupling) specific heat from the experiment[15].

*Specific heat.* An analysis of C(T) below $T_c$ in the framework of the $\alpha$-model[15] with $\alpha$=1.776, within 1% of the BCS value. C/T renormalization is 1.6, to be compared to the calculated [6][7] $\lambda$=0.85. This seems like a good agreement, but, as Fig. 5 shows, the calculated $\alpha^2F(\omega)$ produces strong coupling effects in C(T) well beyond the experimental error. We have also checked that the disagreement is stable against the shape of $\alpha^2F(\omega)$, by repeating the calculations using the same $\lambda$=0.85, but condensing the $\alpha^2F(\omega)$ to one Einstein mode. The shape of $\Delta C(T)/T$ did change, but not enough to explain the experiment.

*Isotope effect.* Hinks *et al*[20] have measured the isotope effect on Ca and found it to be 0.5, as opposed to 0.25 predicted by the calculations[6]. Since all Ca modes are rather soft, coupling exclusively with Ca phonons must necessarily be very strong, in disagreement with the specific heat data suggesting a typical weak coupling regime, and with other indications of weak coupling.

Thus, the main challenge regarding the superconductivity in $AC_6$ is to reconcile the evidence for weak coupling (i.e., dominance of high-energy C modes)[15] with evidence for the dominance of the intercalant modes (i.e., strong coupling)[20] and with the computations[6][7] and experiment[21] (this experiment showed that both types of modes are needed to explain quantitatively the temperature dependence of resistivity, and that with pressure the coupling with all modes remains constant, while the Ca modes soften until becoming unstable) predicting intermediate coupling with both C and Ca equally involved.